\documentclass[twocolumn,showpacs,preprintnumbers,amsmath,amssymb]{revtex4}
\usepackage{graphicx}% Include figure files
\usepackage{dcolumn}% Align table columns on decimal point
\usepackage{bm}% bold math

\begin{document}

%\preprint{APS/123-QED}

\title{The strength of nuclear shell effects at $N=126$ in the r-process 
region}

\author{A.R. Farhan and M.M. Sharma}
 \affiliation{Physics Department, Kuwait University, Kuwait 13060}

\date{\today}% It is always \today, today,
             %  but any date may be explicitly specified

\begin{abstract}
We have investigated nuclear shell effects across the magic number
$N=126$ in the region of the r-process path. Microscopic calculations 
have been performed using the relativistic Hartree-Bogoliubov (RHB) 
approach within the framework of the Relativistic Mean-Field (RMF) 
theory for isotopic chains of rare-earth nuclei in the r-process 
region. The Lagrangian model NL-SV1 with the inclusion of the vector 
self-coupling of $\omega$ meson has been employed. The RMF results show 
that the shell effects at $N=126$ remain strong and exhibit only a slight 
reduction in the strength in going from the r-process path to the neutron 
drip line. This is in striking contrast to a systematic weakening 
of the shell effects at $N=82$ in the r-process region predicted 
earlier in the similar approach. In comparison the shell effects with 
microscopic-macroscopic mass formulae show a near constancy of 
shell gaps leading to strong shell effects in the region of 
r-process path to the drip line. A recent analysis of solar-system 
r-process abundances in a prompt supernova explosion model 
using various mass formulae including the recently introduced mass 
tables based upon Hartree-Fock-Bogoliubov method shows that 
whilst mass formulae with weak shell effects at $N=126$ give rise 
to a spread and an overproduction of nuclides near the third
abundance peak at $A \sim 190$, mass tables with droplet models showing 
stronger shell effects are able to reproduce the abundance features 
near the third peak appropriately. In comparison, several analyses of 
the second r-process peak at $A \sim 130$ have required weakened 
(quenched) shell effects at $N=82$. Our predictions in the RMF theory 
with NL-SV1, which exhibit weaker shell effects at $N=82$ and correspondingly
stronger shell effects at $N=126$ in the r-process region, support the 
conjecture that a different nature of the shell effects at the magic 
numbers may be at play in r-process nucleosynthesis of heavy nuclei.

\end{abstract}

\pacs{~21.10.Dr, 21.30.-x, 21.60.-n, 25.30.+k, 26.50.+x}% PACS, the Physics and Astronomy
                             % Classification Scheme.
%\keywords{Suggested keywords}%Use  showkeys class option if keyword
                              %display desired
\maketitle

\section{Introduction}
\label{intro}
About half of nuclei heavier than Fe are synthesized in the process
of rapid neutron capture (r-process) 
\cite{BBFH.57,Hille.78,Cowan.91,Kratz.93,Kratz.00}. 
In environments of high neutron densities and high temperatures, 
extremely neutron-rich nuclei with at least
10-30 neutrons away from the stability line are produced.
These nuclei are highly unstable and experimentally 
inaccessible, especially those in the heavy mass region. 
The ensuing nuclei undergo a sequence of neutron capture 
accompanied by a spate of $\beta^-$ decays thus
leading to formation of heavy elements in nature. 
The r-process path passes through the magic numbers $N=$~50, 82 and 
126 at different mass values. The synthesis of nuclei around these 
magic numbers is reflected vividly in the known nuclear abundance
peaks around  $A \sim$ 80, 130 and 190, respectively. 

The shell effects at the magic numbers play a significant role in
determining the r-process nuclear abundances \cite{Kratz.93}.
The question whether the shell effects near the r-process 
path are strong or do quench has become crucial to understanding 
the nucleosynthesis of heavy nuclei \cite{Pfeiffer.01}.
This question remains open to-date and the existing data do not suffice
to answer this question. This is due to the reason that r-process 
nuclei are extremely neutron rich and are not accessible experimentally. 
Moreover, on the basis of a few nuclei that are known in the 
extreme regions, it is not easy to make reliable predictions  in
the farther regions of the period table. Consequently, it is proving 
to be difficult to ascertain the nature of the shell effects 
in the vicinity of the r-process path. A knowledge about 
properties of these nuclei is, therefore, obtained from theoretical 
models. At the same time, new data in unknown regions are being
obtained experimentally. Such data can be of enormous value in
defining the nature of properties of nuclei in the extreme regions.

In principle, microscopic calculations within a reliable model 
would be attractive for the purpose. The primary condition on utility
of a microscopic framework should be its ability to reproduce features
and properties of nuclei in the known domains with better accuracy.
Calculations within a microscopic model for a considerably large 
number of nuclei can be cumbersome. However, with the progress in 
computing speeds, such a task is no more beyond one's reach. 

Heretofore, macroscopic-microscopic approaches have largely been used to 
calculate and extrapolate properties of nuclei in the inaccessible 
regions. Most prominent amongst these is the approach of the 
Finite-Range Droplet Model (FRDM) \cite{Moeller.95}. The mass formula 
FRDM has been obtained on the basis of extensive fits of more than 
a thousand known nuclei across the period table including those 
discovered at the periphery of the periodic table in the last decade. 
R-process calculations have been performed using the
binding energies (masses) and neutron separation energies from 
the FRDM in conjunction with $\beta$-decay properties obtained in the 
Quasi-Particle Random Phase Approximation (QRPA) \cite{Moeller.90}. 
Another mass formula that has also been employed extensively is based 
upon the Extended Thomas-Fermi model with Strutinsky Integral (ETF-SI) 
\cite{Abou.95}. Herein, the liquid drop (or the smooth) part is provided 
by the Extended Thomas-Fermi model and the shell corrections are 
superimposed thereupon using the method of the Strutinsky shell 
correction \cite{Stru.68}.

Employing the results of these two mass formulae extensive r-process
network chain calculations have been undertaken
\cite{Kratz.93,Kratz.00,Pfeiffer.01}. It was concluded that
due to strong shell effects (gaps) at $N=82$ and at $N=126$ in 
the region of the r-process path inherent in the mass models 
FRDM and ETF-SI, strong deficiencies (troughs) are obtained in 
reproducing solar system r-process abundances below the peaks 
$A \sim 130$ and $A\sim 190$ \cite{Kratz.93,Pfeiffer.01}.
A remedial measure was suggested by the Hartree-Fock-Bogoliubov 
(HFB) calculations \cite{Doba.96} using the Skyrme force SkP. 
This force is shown to quench (weaken) the strength of the
shell effects in the r-process region significantly, especially at $N=82$. 
This feature of HFB+SkP has proved to be useful in filling up the troughs
below $A \sim 130$ and $A\sim 190$ in the r-process abundance curve.
In this approach, the quenching effect is attributed mainly to a large
effective mass $m*=1$ of the force SkP. Inspired by the usefulness of
the quenching along the r-process path, contrary to the original feature
of both the FRDM and the ETF-SI, quenching has been introduced in the 
new variant of the mass formula, viz., ETF-SI~(Q) \cite{Pear.96}. 
The mass table ETF-SI~(Q) has been shown to be successful in removing 
the main deficiencies in the abundance curve \cite{Pfeiffer.97}.
These experiments with the mass formulae have indicated the need of
weaker shell effects along the r-process path.  A microscopic basis 
of such a requirement at the r-process path, however, needs to be
examined.

In a recent investigation of the r-process nucleosynthesis of heavy nuclei
using mass formulae based upon Hartree-Fock-Bogoliubov approach, it has been
shown \cite{Wanajo.04} that weak shell effects in microscopic mass formulae
result in a spread of abundance distribution near the $A \sim 130$ 
and $A \sim 190$ peaks. This has the consequence that large deviations are
observed compared to the solar-system abundances especially for the peak
about $A \sim 190$. In an earlier analysis \cite{Thielemann.94}, it
was also shown that in a realistic astrophysical scenario a mass model
without quenching at $N=126$ can fill up deficiencies (troughs) 
near $A \sim 175$ due to freeze-out effects. However, this seems to apply
to only the third peak in the abundance curve. Thus, the role of the shell
effects at $N=126$ in the r-process region is not yet clear. 
 
In our earlier work \cite{Sharma.02,Sharma.01,Farhan.03}, we investigated 
the behaviour of the shell effects at the magic number $N=82$ within
the Relativistic Hartree-Bogoliubov approach. Using the Lagrangian model 
of the nonlinear quartic coupling of $\omega$ meson in the RMF theory, 
it was shown that microscopic RMF calculations show a weakening 
of the shell effects at $N=82$ in going from the stability line to the 
r-process path. It was also shown that in going from the r-process 
path to the neutron drip line, the the shell gap diminishes 
to a vanishingly low value at a given isospin, resulting in a complete 
washout of the shell effects \cite{Sharma.02}. In the present work, 
we have investigated the behaviour and evolution of the shell effects 
at $N=126$ in the region of the r-process path within the framework of 
the RHB approach. The salient features of the formalism are discussed 
in Section~\ref{rmf} with emphasis on the shell properties of nuclei 
as inherent to the subject. Results of microscopic RMF calculations 
using two different Lagrangian models are presented in 
Section~\ref{results}. A comparison of the results is made with the 
predictions of various macroscopic-microscopic mass formulae and
influence of shell effects on the r-process nucleosynthesis is 
discussed. A discussion is presented of the possible consequences of
the shell effects on the r-process nucleosynthesis. A summary of the 
results is presented in the last section.

\section{The shell effects in nuclei}
\label{shell}

The shell effects constitute an important feature of nuclei and 
are known to manifest strongly in terms of the magic numbers. 
This is signified by a conspicuous presence of prominent kinks 
about the major magic numbers in two-neutron separation 
energies ($S_{2n}$) all along the stability line \cite{Bor.93}. 
This is a manifestation of the existence of large shell gaps at 
magic numbers in nuclei. The spin-orbit interaction plays a 
pivotal role in the creation of the shell gaps and consequently
of the magic numbers. Of late, there are indications that with extreme
isospin in light nuclei, re-adjustment of single-particle levels might 
lead to an emergence of new magic numbers \cite{Fridmann.05} other than those 
which are are hitherto established.  

The spin-orbit interaction and consequently how the shell effects 
behave in the extreme regions would play a significant role in carving 
out shell gaps in nuclei near the r-process path. 
In microscopic approaches such as nonrelativistic 
density-dependent Skyrme theory and the relativistic mean-field 
theory, the spin-orbit interaction is determined 
by data on a few nuclei. Whilst in the former, the spin-orbit 
interaction is added in the theory on an {\it ad hoc} basis and its 
strength is adjusted to known spin-orbit splittings in a few nuclei, 
it arises naturally in the RMF theory as a consequence of the 
Dirac-Lorentz structure of nucleons. This has shown much usefulness 
in explaining properties that involve shell effects, such as anomalous 
isotope shifts in stable nuclei, especially those associated to
the Pb chain \cite{SLR.93}. This feature of the shell effects has not 
been possible to attain in the Skyrme theory without undertaking 
significant alterations in the isospin dependence of the spin-orbit 
interaction \cite{RF.95}. On the other hand, the intrinsic form of the 
spin-orbit interaction in the RMF theory has been found to be 
advantageous over that in the nonrelativistic approach. The appropriate
isospin dependence of the spin-orbit interaction \cite{SLK.94}
has been found to be successful in reproducing the anomalous 
isotope shifts in Pb nuclei \cite{SLR.93} as well as in Sr and Kr 
isotopes \cite{LS.95}. Consequently, it is expected to have 
implications in predicting the shell strength in the extreme
regions of the r-process path.

The RMF theory \cite{SW.86,Rein.89,Ser.92,Ring.96} has shown an immense 
potential in being able to describe properties of nuclei along the 
stability line \cite{GRT.90,SNR.93} and for a large number 
of nuclei beyond the stability line. Most of the 
Lagrangian parameter sets are based upon reproduction of binding 
energies, charge radii and in some cases surface thickness of a 
few key nuclei \cite{Rein.89}. Various forces are obtained 
in such a way that spin-orbit splitting in some key nuclei such 
as $^{16}$O is reproduced reasonably well. It should, however, be 
pointed out that it does not necessarily ensure that shell gaps or 
shell effects at the major magic numbers are reproduced correctly. 

The shell effects were not considered explicitly in the initial 
developments in the RMF theory of finite nuclei. This problem was 
addressed in Ref. \cite{Sharma.00}, where shell effects were investigated in 
nuclei along the stability line with a view to see their influence on 
nuclei near r-process path or on some known ``waiting-point'' 
nuclei \cite{Sharma.02}. It was shown \cite{Sharma.00} that the otherwise 
successful RMF forces based upon non-linear self-coupling of $\sigma$-meson 
such as NL-SH \cite{SNR.93} overestimate the experimental shell gaps in 
nuclei along the stability line. In order to solve this problem, 
the Lagrangian model with the nonlinear scalar coupling of $\sigma$ meson
was extended with the inclusion of the nonlinear quartic coupling of 
the $\omega$ meson \cite{Sharma.00}. Consequently, shell effects in 
Ni and Sn isotopes at the stability line were reproduced well.

The shell effects along the stability line (may) have repercussions 
(as it seems to be the case for the RMF theory, but not 
unequivocally for the mass formulae) on the shell effects far 
away from it. The character of the shell effects, be it
strong or weak vis-a-vis experimental data along the line of stability 
is likely to extrapolate alike (unless major re-adjustments in 
single-particle scheme take place giving rise to unexpected pattern of
behaviour) in the unknown regions of the periodic 
table. A test case for this hypothesis was provided by the waiting-point 
nucleus $^{80}$Zn ($N=50$) which lies close to the r-process path at $N=50$.
It was shown \cite{Sharma.02} that forces such as NL-SH that 
overestimate the shell effects at the stability line overestimate the 
shell effects for the waiting-point nucleus $^{80}$Zn and also in r-process
nuclei at $N=82$. On the other hand, the force NL-SV1 based upon the 
vector self-coupling of $\omega$-meson, which reproduces the shell 
effects in nuclei at the stability line, is able to reproduce the available 
data on the waiting-point nucleus $^{80}$Zn \cite{Sharma.02}. A firmer
verification of predictability of various theories would be provided
by future experimental data in the extreme regions. In the present work,
we explore how this feature translates for the shell 
effects at $N=126$ in the r-process region.

\section{The Relativistic Mean-Field theory}
\label{rmf}

The RMF approach \cite{SW.86} is based upon the Lagrangian density 
which consists of fields due to various mesons interacting with 
nucleons. The mesons include the isoscalar scalar $\sigma$-meson, 
the isoscalar vector $\omega$-meson and the isovector vector $\rho$-meson.
The details of the formalism can be found in 
Refs. \cite{Sharma.02,Rein.89,Ring.96,GRT.90}.

The RMF Lagrangian that describes the nucleons as Dirac spinors 
moving in the meson fields is given by \cite{SW.86}

\begin{eqnarray}
{\cal L}&=& \bar\psi \left( \rlap{/}p - g_\omega\rlap{/}\omega -
g_\rho\rlap{/}\vec\rho\vec\tau - \frac{1}{2}e(1 - \tau_3)\rlap{\,/}A -
g_\sigma\sigma - M_N\right)\psi\nonumber\\
&&+\frac{1}{2}\partial_\mu\sigma\partial^\mu\sigma-U(\sigma)
-\frac{1}{4}\Omega_{\mu\nu}\Omega^{\mu\nu}+ \frac{1}{2}
m^2_\omega\omega_\mu\omega^\mu\\ &&+\frac{1}{4}g_4(\omega_\mu\omega^\mu)^2
-\frac{1}{4}\vec R_{\mu\nu}\vec R^{\mu\nu}+
\frac{1}{2} m^2_\rho\vec\rho_\mu\vec\rho^\mu -\frac{1}{4}F_{\mu\nu}F^{\mu\nu}
\nonumber
\end{eqnarray}
where $M_N$ is the bare nucleon mass and $\psi$ is its Dirac spinor. 
Nucleons interact with $\sigma$, $\omega$, and $\vec\rho$) mesons, 
with the masses being $m_\sigma$, $m_\omega$ and $m_\rho$ and the 
coupling constants being $g_\sigma$, $g_\omega$, $g_\rho$, respectively. 
The electromagetic interaction is
represented by the electromagnetic vector field $A^\mu$. 

The field tensors for the vector mesons are given as 
$\Omega_{\mu\nu}=\partial_\mu\omega_\nu-\partial_\nu\omega_\mu$ 
and by similar expressions for the $\rho$-meson and 
the photon. For a realistic description of nuclear properties a nonlinear
self-coupling 
\begin{eqnarray}
U(\sigma) = \frac{1}{2} m^2_\sigma \sigma^2_{} + \frac{1}{3}g_2\sigma^3_{} 
+ \frac{1}{4}g_3\sigma^4 
\end{eqnarray}
for $\sigma$-mesons has been widely used. The non-linear vector self-coupling 
of $\omega$-meson \cite{Bod.91} as added earlier \cite{Sharma.00} in 
the Lagrangian with the non-linear scalar field is represented by the 
coupling constant g$_4$. 

In the lowest order of the quantum field theory, i.e., in the  mean-field 
approximation, the nucleons are assumed to move independently in the 
meson fields. The meson fields are replaced by their classical expectation 
values. The variational principle leads to the Dirac equation:
\begin{eqnarray}
\{ -i{\bf {\alpha}} \nabla + V({\bf r}) + \beta {m*} \}
~\psi_{i} = ~\epsilon_{i} \psi_{i}
\end{eqnarray}
where $V({\bf r})$ represents the $vector$ potential:
\begin{eqnarray}
V({\bf r}) = g_{\omega} \omega_{0}({\bf r}) + g_{\rho}\tau_{3} {\bf {\rho}}
_{0}({\bf r}) + \frac{e(1-\tau_{3})}{2} {A}_{0}({\bf r})
\end{eqnarray}
and $S({\bf r})$ is the $scalar$ potential
\begin{eqnarray}
S({\bf r}) = g_{\sigma} \sigma({\bf r})
\end{eqnarray}
which defines the effective mass as:
\begin{eqnarray}
m^{\ast}({\bf r}) = m + S({\bf r})
\end{eqnarray}
The Klein-Gordon equations for the meson fields are time-independent
inhomogeneous equations with the nucleon densities as sources.
\begin{eqnarray}
\{ -\Delta + m_{\sigma}^{2} \}\sigma({\bf r})
 &=&-g_{\sigma}\rho_{s}({\bf r})
-g_{2}\sigma^{2}({\bf r})-g_{3}\sigma^{3}({\bf r})
\nonumber \\ 
\  \{ -\Delta + m_{\omega}^{2} \} \omega({\bf r})
&=& g_{\omega}\rho_{v}({\bf r}) + g_4 \omega^3({\bf r}) 
\nonumber \\ 
\  \{ -\Delta + m_{\rho}^{2} \}\rho({\bf r})
&=& g_{\rho} \rho_{3}({\bf r})
\nonumber \\ 
\  -\Delta A({\bf r}) = e\rho_{c}({\bf r})
\end{eqnarray}

The stationary state solutions $\psi_i$ are obtained from the coupled 
system of Dirac and Klein-Gordon equations. The ground-state of the 
nucleus is described by a Slater determinant $\vert\Phi >$ of single-particle
spinors $\psi_i$ (i = 1,2,....A). Solution of the Dirac equation 
is achieved by using the method of oscillator expansion \cite{GRT.90}. 
In the RMF approach, the pairing is included within the BCS scheme.
However, for the case of nuclei in the extreme regions of the r-process
path and drip lines, the Fermi energy is very close to the continuum and
many single-particle states couple to the continuum. Thus, the BCS method
of pairing provides a crude approximation of such cases. The Relativistic 
Hartree-Bogoliubov (RHB) approach based upon quasi-particle scheme provides 
an appropriate framework to deal with nuclei of such a nature.

\subsection{The Relativistic Hartree-Bogoliubov approach}

Nuclei that are known to show strong pairing correlations
are treated appropriately within the framework of the
RHB approach. Pairing correlations in the neighbourhood of 
the Fermi energy in case of nuclei near r-process path and drip 
line become even more important. Herein, the RHB approach provides
a suitable framework to deal with nuclei in the extreme regions.

It has been shown \cite{KR.91} that using Green's function 
techniques \cite{Go.58} a relativistic Hartree-Bogoliubov 
approach can be implemented using the Lagrangian as given above. 
Neglecting retardation effects and the Fock term, one obtains  
relativistic Dirac-Hartree-Bogoliubov (RHB) equations
\begin{equation}
\left(\begin{array}{cc} h & \Delta \\ -\Delta^* & -h^* \end{array}\right)
\left(\begin{array}{r} U \\ V\end{array}\right)_k~=~
E_k\,\left(\begin{array}{r} U \\ V\end{array}\right)_k,
\label{RHB} 
\end{equation}
where $E_k$ are quasiparticle energies and the coefficients $U_k$ and 
$V_k$ are four-dimensional Dirac spinors normalized as
\begin{equation}
\int ( U^+_k U^{}_{k'}~+~V^+_kV^{}_{k'}\, ) d^3r~=~\delta_{kk'}.
\end{equation} 
The average field
\begin{equation}
h~=~\mbox{\boldmath $\alpha.p$}~+~g_\omega\omega~+~ \beta(M+g_\sigma
\sigma)~-~\lambda
\label{h-field}
\end{equation}
contains the chemical potential $\lambda$. 
The meson fields $\sigma$ and $\omega$ are determined self-consistently 
from the Klein Gordon equations as done in the case of the 
RMF equations discussed above with the scalar density 
$\rho_s=\sum_k \bar V^{}_kV^{}_k$ and the baryon density 
$\rho_v=\sum_k V^+_kV^{}_k$.  The sum on $k$ is taken only over 
the particle states in the no-sea approximation. 
The pairing potential 
$\Delta$ in Eq. (\ref{RHB}) is given by
\begin{equation}
\Delta_{ab}~=~\frac{1}{2}\sum_{cd} V^{pp}_{abcd} \kappa_{cd}
\label{pair}
\end{equation}
The RHB equations (\ref{RHB}) are a set of four coupled 
integro-differential equations for the Dirac spinors $U(r)$ and 
$V(r)$ that are obtained self-consistently. The RHB calculations 
are performed by expanding fermionic and bosonic wavefunctions 
in 20 oscillator shells. For the pairing channel, we use the 
finite-range Gogny force D1S \cite{Berger.84}. The Gogny force is 
a sum of two Gaussians with finite range. It has been shown 
\cite{Berger.84} that the Gogny force is able to describe 
pairing properties of a large number of finite nuclei in the
medium and heavy mass regions. Details of the RHB theory can be 
found in Ref. \cite{KR.91}

\subsection{Lagrangian Models}

The Lagrangian model with the nonlinear scalar coupling of $\sigma$ meson
has been the widely used one for finite nuclei within the RMF theory. 
It has been successful in reproducing ground-state properties of 
nuclei at the stability line as well as of those far away from it. 
Here, we will consider the successful forces NL-SH \cite{SNR.93} 
and NL3 \cite{Lala.97} within this Lagrangian model. 
We will also employ the forces NL-SV1 and NL-SV2 \cite{Sharma.05} 
(see ref. \cite{Sharma.00} for the parameter sets) with the 
nonlinear vector self-coupling of $\omega$ meson. 
As mentioned above, forces NL-SV1 and NL-SV2 were constructed with the 
inclusion of the quartic vector coupling of $\omega$ meson, in order 
to solve the problem of strong shell effects with Lagrangian model with 
nonlinear scalar coupling of $\sigma$ meson \cite{Sharma.00}. 
The introduction of the non-linear coupling of $\omega$-meson also
softens the equation of state (EOS) of the nuclear matter significantly. 
This has the consequence that the maximum neutron star mass with such 
an EOS would show a better agreement with empirically observed 
values. A detailed discussion of the properties associated with the 
introduction of the nonlinear vector self-coupling of $\omega$ meson 
in the RMF theory will be presented elsewhere \cite{Sharma.05}. It will 
be shown \cite{Sharma.05} that the Lagrangian parameter set NL-SV1 is 
also able to improve upon the ground-state properties such as binding 
energies, charge radii and isotopes shifts of nuclei along the stability
line and far away from it as compared to those with NL-SH and NL3.

\section{Details of the calculations}

In RHB calculations, wavefunctions are expanded into an
harmonic-oscillator basis to solve the Dirac and the Klein-Gordan 
equations \cite{GRT.90}. For both the fermionic and bosonic fields 
a basis of 20 oscillator shells has been used. The pairing has 
been taken in the Bogoliubov approach. For, nuclei a few neutrons 
below and a few neutrons above a magic number are usually 
spherical, RHB calculations have been performed for a 
spherical configuration. For a comparative study of the shell 
effects, we have performed RHB calculations with the two Lagrangian 
models as discussed above. However, our focus is on investigation of 
the potential and predictive power of the new Lagrangian model with the 
vector self-coupling of $\omega$ meson vis-a-vis the scalar 
self-coupling of $\sigma$ meson.

With a view to investigate as to how the shell effects evolve in going
from the region of the stability line towards the r-process path and 
ultimately to the neutron drip line, we have selected even-even nuclei 
from the isotopic chains of Hf ($Z=72$) down to Xe ($Z=54$) across 
the neutron magic number $N=126$.  For our focus is on the behaviour of
the shell gap, nuclei relevant to the discussion are those with $N=124$, 
$N=126$ and $N=128$, so as to be able to calculate $S_{2n}$ values 
across the magic number $N=126$.

\section{Results and Discussion}
\label{results}
\subsection{Shell effects along the stability line}
First, we examine the known shell effects at $N=126$ at the stability line.
The experimental shell gap at $N=126$ is known only in a very few 
isotopic chains in this region. Here, the doubly closed nucleus 
$^{208}$Pb in the Pb chain provides a cardinal point in the study 
of the shell effects at $N=126$. 
Interestingly, though the nuclear landscape of the periodic table 
has been extended significantly due to sustained experimental 
efforts in the last decade towards synthesizing nuclei far beyond the 
stability line in the laboratory, the heaviest known Pb nucleus 
has reached only to $^{214}$Pb. 

We begin with the premise that the two-neutron separation energy 
at the magic number provides a reasonably good indicator of the shell gap.
Therefore, we calculate the shell gap at the magic number as
defined conveniently by
\begin{equation}
\Delta_S = S_{2n}(Z, N_0) - S_{2n}(Z, N_0+2),
\label{gap}
\end{equation}
where $S_{2n}(Z, N_0)$ denotes the 2-neutron separation energy of the
nucleus $(Z, N_0)$ with a magic neutron number $N_0$.

The shell gap in Pb nuclei obtained from the RMF approach using various
Lagrangian sets is shown in Fig.~1(a) and (b). The corresponding gap 
from various macroscopic-microscopic mass formulae is shown in Fig.~1(c).
The results are compared with the experimentally known data. 
The $S_{2n}$ values calculated in the RMF theory with the forces 
NL-SH and NL3 within Lagrangian model of the nonlinear scalar 
coupling of $\sigma$ meson are shown in Fig.~1(a). 
The difference between the data points at $N=126$ and $N=128$, 
as indicated by Eq.~\ref{gap}, manifests the shell gap at the
magic number $N=126$. A comparison with the experimental data points shows
that the shell gap from the force NL-SH underestimates the experimental
gap by $\sim 0.5$ MeV. On the other hand, the recent force NL3 shows a good 
agreement with the data. Here, we do not show the binding energy of 
$^{208}$Pb itself that is overestimated both by NL-SH and NL3 by 
about 2-3 MeV. However, the difference in the $S_{2n}$ values of the
neighboring nuclei turns out to be satisfactory in both the cases.
It is interesting to note that although NL-SH underestimates 
the shell gap at $N=126$ slightly, it is known to exhibit 
generally stronger shell effects for nuclei at and beyond the
stability line than those from NL3. This implies that a focus on a 
single or a few data points may not be deterministic as far as 
extrapolations (predictions) for nuclei beyond the stability line
are concerned. Consequently, the shell gap at $N=126$ in the Pb 
nucleus may not provide a successful conjecture as to whether the 
shell effects with either of the forces shall remain strong or 
weak in the domain that is far beyond the stability line.

% For one-column wide figures use
\begin{figure}
% Use the relevant command for your figure-insertion program
% to insert the figure file.
% For example, with the option graphics use
%\centering
\hspace{-0.5cm}
\resizebox{0.45\textwidth}{!}{%
  \rotatebox{270}{\includegraphics{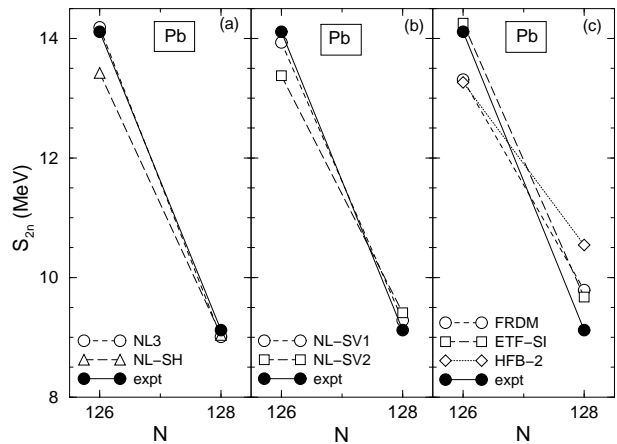}}
}
% If not, use
\vspace{0cm}       % Give the correct figure height in cm
\caption{The two-neutron separation energies $S_{2n}$ for Pb isotopes
as obtained from (a) the forces with the Lagrangian model of the nonlinear
scalar self-coupling of $\sigma$ meson, (b) from the forces which
include the nonlinear quartic coupling of $\omega$ meson, and (c)
from the mass formulae FRDM \cite{Moeller.95}, ETF-SI \cite{Abou.95} 
and the recently obtained results from mass tables HFB-2 \cite{Goriely.02}.
The experimental data is shown by the solid circles.}
\label{fig:1}       % Give a unique label
\end{figure}

% For one-column wide figures use
\begin{figure*}
% Use the relevant command for your figure-insertion program
% to insert the figure file.
% For example, with the option graphics use
\centering
\resizebox{0.65\textwidth}{!}{%
  \rotatebox{270}{\includegraphics{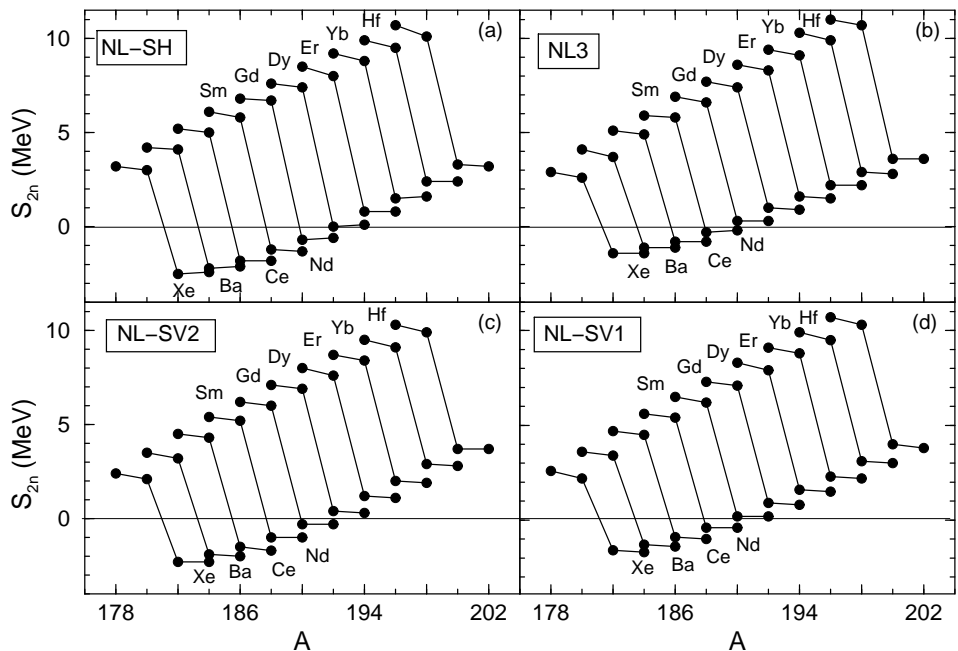}}
}
% If not, use
\vspace{1cm}       % Give the correct figure height in cm
\caption{The two-neutron separation energies $S_{2n}$ for the isotopic
chains from Hf ($Z=72$) to Xe ($Z=54$) in going towards the r-process nuclei
and in approaching the neutron drip line, obtained from the forces with the 
Lagrangian model of the nonlinear scalar self-coupling of 
$\sigma$ meson (a) NL-SH \cite{SNR.93} and (b) NL3 \cite{Lala.97}. 
The $S_{2n}$ values from the forces which include the nonlinear 
quartic coupling of $\omega$ meson \cite{Sharma.00,Sharma.05}
are also shown in (c) with NL-SV2  and in (d) with NL-SV1.} 
\label{fig:2}       % Give a unique label
\end{figure*}

The $S_{2n}$ values obtained from the forces NL-SV1 and NL-SV2 with the 
Lagrangian model with the nonlinear quartic coupling of $\omega$ meson 
are shown in Fig.~1(b). A comparison with the experimental data shows
that NL-SV1 reproduces the shell gap at $N=126$ well, whereas it is
underestimated by $\sim 0.5$ MeV by the force NL-SV2. Again, a
seemingly paradoxical situation arises here. As in the case of NL-SH,
exhibiting generally stronger shell effects and yet underpredicting 
the shell gap at $N=126$ as shown in Fig.~1(a), the force NL-SV2 has 
also been shown to exhibit shell effects slightly stronger than those 
with NL-SV1 in the r-process region at $N=82$ \cite{Sharma.02}. 
This shows that reliance on a few data points may not be useful in 
predicting behaviour in the extreme regions. This will be shown in 
the latter parts of this paper, where we will discuss the shell 
effects in the region of the r-process path and the neutron drip line.

It is equally interesting to see as to how various mass formulae predict 
the shell gap at $N=126$ in Pb nuclei at the stability line.
In Fig.~1(c), the data points from the mass formula FRDM and ETF-SI
are shown. It may be remarked that various mass formulae including 
FRDM and ETF-SI have been obtained with a view to reproducing 
experimental data on more than a thousand nuclei. Generally, these 
mass tables have achieved a great success in reproducing a large set of 
experimental database through exhaustive fits over the periodic table 
than possibly a microscopic theory could ever do. However, as pointed 
out in the literature \cite{Moeller.95,Goriely.02}, 
discrepancies at the magic numbers do remain a significant drawback. 

The FRDM indicates a shell gap that is $\sim 1.5$ MeV 
smaller than the experimental one. On the other hand, ETF-SI 
underestimates the shell gap only by about 0.5 MeV. The undervaluation
of the shell gap with FRDM and ETF-SI at $N=126$ along the 
stability line is to be contrasted to the stronger shell effects 
due to these mass formulae when extrapolated in the extreme regions 
of the r-process path both at $N=82$ and $N=126$, than are suggested
for a successful reproduction of r-process abundances. Notwithstanding 
the need  of weaker shell effects along the r-process path, a new mass table 
based upon the Skyrme Hartree-Fock-Bogoliubov (HFB) approach has 
recently been produced \cite{Goriely.02}. The data points from the mass
table HFB-2 as shown in Fig.~1(c) underestimate the shell gap 
at $N=126$ significantly. It is worth pointing out that the 
mass formula HFB-2 \cite{Goriely.02} seems to have 
achieved a similar quality of fit across much of the periodic table. 
It is comparable to ETF-SI and FRDM, albeit with shell gaps that are 
predicted to be small in the r-process region, as we will see 
in Section~\ref{mass}.

\begin{figure*}
% Use the relevant command for your figure-insertion program
% to insert the figure file.
% For example, with the option graphics use
%\rotatebox{90}
\centering
%\vspace{0.5cm}
\resizebox{0.65\textwidth}{!}{%
  \rotatebox{270}{\includegraphics{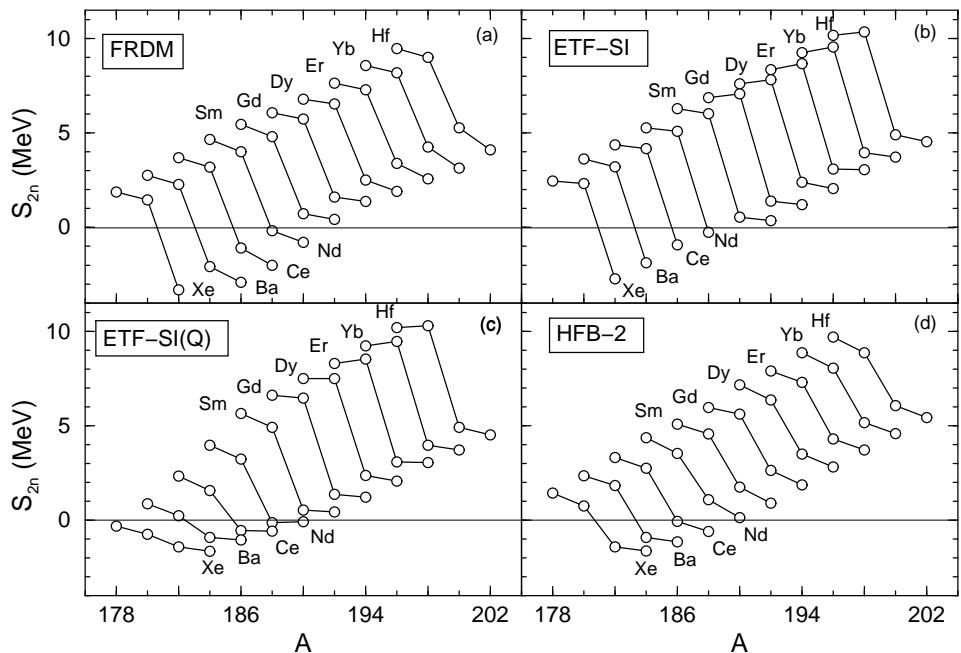}}
}
% If not, use
\vspace{1cm}       % Give the correct figure height in cm
\caption{The two-neutron separation energies $S_{2n}$ for the isotopic
chains from Hf ($Z=72$) to Xe ($Z=54$) in going towards the r-process nuclei
and in approaching the drip line, as predicted by the mass models (a) FRDM 
\cite{Moeller.95} and (b) ETF-SI \cite{Abou.95}. The $S_{2n}$ values 
from the mass formula ETF-SI~(Q) \cite{Pear.96} that includes a quenching 
of the shell effects superimposed on ETF-SI are shown in (c). The results 
from the mass formula HFB-2 \cite{Goriely.02} are shown in (d).}
\label{fig:3}       % Give a unique label
\end{figure*}

\subsection{Shell effects near the r-process path - RMF theory}

The shell effects at $N=126$ for r-process nuclei play a crucial role in
determining r-process abundances around the peak at $A \sim 190$ 
\cite{Kratz.93}. For practical purposes r-process nuclei are 
defined to be those with $S_n \sim 2-4$ MeV. Thus, the r-process path 
is not strictly well defined and it does vary from model to model. 
However, it is generally accepted that nuclei with 
$Z \sim 64-69$ near $N=126$ fall along the r-process path.

The results of RHB calculations for the forces NL-SH and NL3 with the
Lagrangian model of nonlinear $\sigma$ coupling are shown in the upper 
panels (a) and (b) of Fig.~2, respectively. It is seen clearly that the
shell gap that is represented by the difference between the $S_{2n}$
values at $N=126$ and at $N=128$ shows a gradual decrease 
with an increase in isospin i.e., in going from the element 
Hf ($Z=72$) that is slightly above the r-process path to Xe ($Z=54$) 
that is near the drip line. This behaviour is similar for both
NL-SH and NL3. In absolute terms the shell gaps for NL-SH are 
slightly larger than those of NL3. However, in both the cases, 
the shell gap does not show a significant decrease in approaching 
the drip line, as is probably expected from a comparison with the 
corresponding behaviour of the shell gap at $N=82$ at the drip line 
\cite{Sharma.02}.

The $S_{2n}$ values obtained from the forces NL-SV1 and NL-SV2 with
the inclusion of quartic coupling of $\omega$ meson are shown in
the lower panels (c) and (d) of Fig.~2, respectively. For both the forces
NL-SV1 and NL-SV2, the shell gaps show a similar gradual decrease in going
towards the r-process path and the drip line. Qualitatively, the 
behaviour of both the Lagrangian models as portrayed in the upper
and the lower panels, respectively, is very similar. However, 
quantitatively, the shell gaps with NL-SV1 are smaller than both with
NL-SH and NL3. The force NL-SV2 provides slightly larger shell gaps than 
those  NL-SV1.

All the parameters sets of both the Lagrangian models exhibit a slight
reduction in the shell strength in the region of the r-process path.
The difference lies only in the degree by which the shell effects are
reduced in going from the r-process path to the drip line. We will discuss
the comparative behaviour of the shell effects with various RMF models
in Section~\ref{compar}.

\subsection{Shell effects near the r-process path - mass models}
\label{mass}

In the absence of and in essence rather infeasibility at present
of constructing a mass table based purely on microscopic calculations, 
masses from various mass tables based upon macroscopic-microscopic approach 
are used in r-process calculations. Here, we present the results of the two 
most elaborate mass formulae, the FRDM and the ETF-SI. 

We show the $S_{2n}$ values obtained from the FRDM and ETF-SI in the 
upper panels (a) and (b) of Fig.~3, respectively. In contrast to the 
microscopic calculations of Fig.~2, the shell gaps with FRDM and 
ETF-SI show nearly constant values in going from $Z=72$ to $Z=54$. 
On the other hand, a slight increase in the shell gap is visible 
with FRDM in going towards Xe. However, as the $S_{2n}$ values 
from FRDM for the nuclides with $N=128$ become negative below 
Nd $(Z=60)$, the apparent increase in the value of the shell gap for 
nuclei below Nd can be therefore be discounted. 

A constant shell gap is also displayed by ETF-SI as shown in Fig.~3(b).
The magnitude of the shell gap with ETF-SI, is, however, larger than 
that with FRDM by $\sim$ 0.8 MeV. On the lower end, i.e, for 
nuclides with $N=128$, the ETF-SI also shows negative $S_{2n}$ 
values below Nd $(Z=60)$. Thus, the drip line is reached
at $N=128$ for the elements below Nd, the behaviour very similar to 
that of FRDM. Moreover, with ETF-SI the $S_{2n}$ values for 
nuclides with the magic number $(N=126)$ are $\sim 1$ MeV 
larger than those of FRDM. The similar behaviour of the
shell effects with FRDM and ETF-SI and arrival of the drip line at a
similar location is not surprising, for the shell corrections 
superimposed on the smooth part in the two models are based upon 
the same prescription of the Strutinsky shell correction \cite{Stru.68}

\begin{figure*}
\centering
%\hspace{4cm}
\vspace{0.5cm}
\resizebox{0.60\textwidth}{!}{%
  \rotatebox{270}{\includegraphics{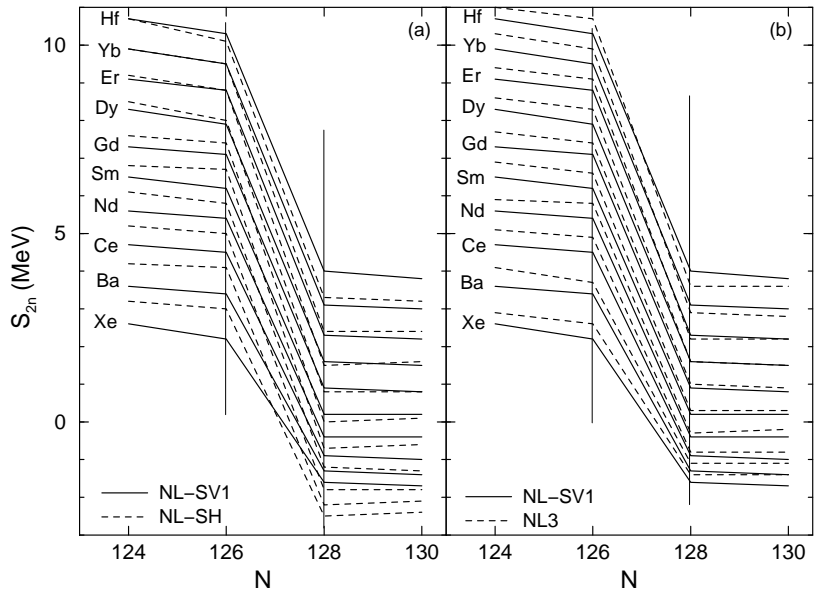}}
}
% If not, use
\vspace{0cm}       % Give the correct figure height in cm
\caption{The $S_{2n}$ values from NL-SV1 are compared to those from
(a) NL-SH and (b) NL3. The shell gap region is bounded by the two vertical
lines at $N=126$ and $N=128$.}
\label{fig:4}       % Give a unique label
\end{figure*}

Predictions from the mass formulae FRDM and ETF-SI in the extreme
regions of the r-process path have been used extensively for network
chain calculations of r-process nuclear abundances. Results of
calculations have shown that due to strong shell effects that are
prevalent  with FRDM and ETF-SI along the r-process path, there is
a significant deficiency (troughs) in the r-process abundances below
the $A \sim 130$ and $A \sim 190$ peaks \cite{Kratz.93}. Inspired by the 
Hartree-Fock-Bogoliubov (HFB) calculations with the Skyrme force
SkP \cite{Doba.96}, quenching was introduced in the ETF-SI mass formula, thus 
creating a new mass table ETF-SI~(Q) \cite{Pear.96}. The resulting 
r-process calculations \cite{Pfeiffer.97} with ETF-SI~(Q) have been 
able to fill up the deficiencies below the two peaks and results 
seem to be promising. This has put up a requirement (indirect)
of a weakening of the shell structure for nuclei near the r-process 
path \cite{Pfeiffer.97,Chen.95}. This feature that is clearly absent 
in the original mass formulae FRDM and ETF-SI has been introduced 
in the ETF-SI (Q) rather artificially. The ETF-SI~(Q) results in 
Fig.~3(c) show that the shell gaps remain the same as that with 
ETF-SI in going from Hf $(Z=72)$ to Er $(Z=68)$. However, the shell gap 
starts decreasing rapidly below Dy $(Z=66)$ in going towards Xe due 
to the quenching introduced therein. A microscopic basis of the 
aforesaid quenching introduced in the mass formula ETF-SI~(Q) 
is yet to be established.

Motivated by the quenching present in the HFB calculations with SkP, attempts
have been made to introduce the HFB approach in some mass tables. We show
in Fig.~3 (d) the results taken from the recently developed mass table 
HFB-2 \cite{Goriely.02} within the Skyrme Ansatz. 
The shell gap with HFB-2 is reduced for all the isotopic 
chains as compared to FRDM and ETF-SI.  A reduction in 
the shell gap with HFB-2 was also seen for Pb isotopes in Fig.~1(c). 
Though the shell gap is reduced vis-a-vis other mass formulae, 
it, however, remains constant in going from Hf $(Z=72)$ to 
Xe $(Z=54)$. There are no indications of an additional reduction 
in the shell strength near the drip line as compared to the region 
of the r-process path. Thus, the constancy of the shell gap
in going from the region of the r-process path to the drip line seems
to be the salient feature of the macroscopic-microscopic mass tables 
presented in Fig.~3. Here, the only exception is ETF-SI (Q), wherein 
the quenching was introduced by force. In comparison, the RMF results show
a slight reduction in the shell strength in going from the r-process path
to the drip line.

An earlier version of the HFB mass table, i.e., HFB-1 was 
constructed by replacing the BCS pairing by the Bogoliubov 
pairing scheme in Ref. \cite{Samyn.02}. In this work, it was shown 
that the behaviour of shell gaps far away from the stability line
does not depend much upon whether the BCS pairing or the Bogoliubov 
pairing is used. Accordingly, the shell gaps at $N=126$ 
with HFB-1 were found to be comparable to those with FRDM and 
that an introduction of Bogoliubov pairing did not result 
in a quenching of the shell effects \cite{Samyn.02}.

\subsection{A comparative analysis of the shell effects}
\label{compar}
In order to understand the comparative behaviour of the shell 
effects first in the RMF theory, we show in Fig.~4 the $S_{2n}$ 
values across the magic number $N=126$ from the force NL-SV1 and 
compare these with the other forces such as NL-SH and NL3. 
It may be reminded that although the force NL-SH has been 
found to be successful in reproducing binding energies, 
charge radii and deformation properties of a large number of
nuclei far away from the stability line, including the anomalous
isotope shifts in Pb nuclei, the shell effects with NL-SH were found
to be stronger as compared to the experimental data \cite{Sharma.00}.
In comparison, the force NL-SV1, having been able to describe the shell 
effects in Ni and Sn isotopes at the stability line, was also shown 
to be successful in reproducing the available data on the shell 
effects at the waiting-point nucleus $^{80}$Zn \cite{Sharma.02}. 
In view of this, we treat the force NL-SV1 as our benchmark. 

\begin{figure*}
\centering
%\hspace{4cm}
\vspace{0.5cm}
\resizebox{0.60\textwidth}{!}{%
  \rotatebox{270}{\includegraphics{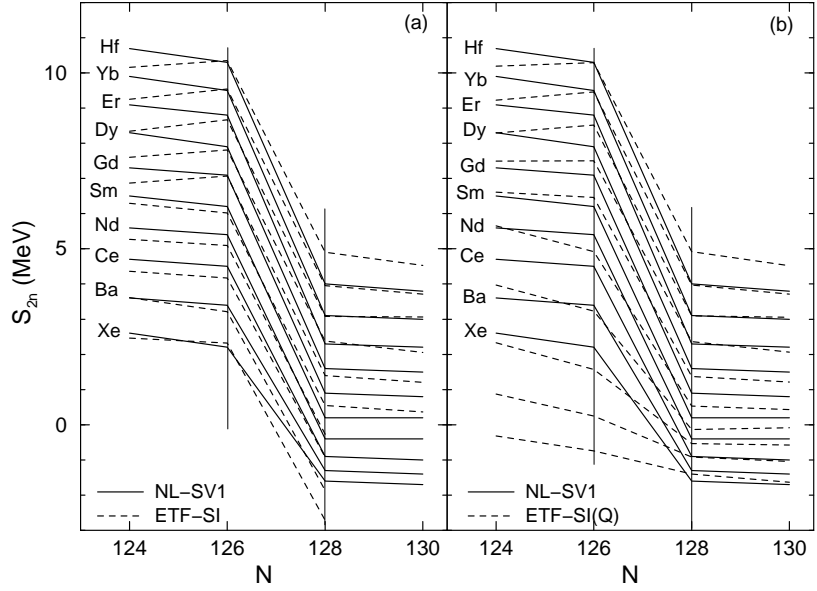}}
}
% If not, use
\vspace{0cm}       % Give the correct figure height in cm
\caption{The $S_{2n}$ values from NL-SV1 are compared to those from the
mass formula (a) ETF-SI and with the quenched mass formula (b) ETF-SI~(Q).
The shell gap region is bounded by the two vertical lines at $N=126$ and 
$N=128$.}
\label{fig:5}       % Give a unique label
\end{figure*}

In Fig.~4(a), we compare the results of NL-SV1 to those from NL-SH. 
The variation in the shell gap is illustrated by the changing slopes
of the curves between the vertical bars at $N=126$ and $N=128$. 
A look at the difference in the $S_{2n}$ values and at the 
corresponding slope of the curve for Hf $(Z=72)$ shows that the 
shell gap with NL-SH is bigger than with NL-SV1.
As one progresses towards the r-process nuclei such as Er, Dy, Gd and
Sm, this difference in the shell gap between NL-SH and NL-SV1
increases with an increase in the isospin. One sees that the shell 
gaps with NL-SH remain stronger even in nuclei near the drip 
line such as Xe. In comparison, NL-SV1 shows a faster decrease in 
the shell gap in going from r-process nuclei to the drip line, a 
feature that has been called for for reproduction of r-process 
nuclear abundances.

The results from NL-SV1 are compared with those from the force
NL3 in Fig.~4(b). A comparison between the shell gaps 
from the two forces shows that beginning with Hf, 
the shell gap with NL3 is slightly larger than that with NL-SV1. 
This difference, however, increases slowly when one moves
from Hf to Xe. Thus, the shell effects with NL3 are slightly stronger
than those with NL-SV1. Additionally, the $S_{2n}$ values for nuclides with
$N=124$ are $\sim 0.5$ MeV higher with NL3 than NL-SV1, especially in the
region of r-process nuclei. This feature is similar to that of NL-SH 
vis-a-vis NL-SV1 as shown in Fig.~4(a). 
We compare in Fig.~5 the shell gaps from NL-SV1 to those from the
mass models (a) ETF-SI and from its variant (b) ETF-SI (Q) that embeds 
a quenching near the r-process region. For Hf ($Z=72$), the shell gap
with NL-SV1 is $\sim 1$ MeV larger than that of ETF-SI. However, as
one proceeds to the r-process nuclei and towards the drip line, 
this difference between NL-SV1 and ETF-SI decreases 
and then it reverses. For drip line nuclei near Xe ($Z=54$), 
the NL-SV1 shell gap is then smaller than that with ETF-SI. 
Whereas the shell gap with ETF-SI hardly shows any change in 
going from Hf to Xe, the shell gap with NL-SV1 does show a consistent 
decrease in going from the r-process to the drip line.

We also compare the NL-SV1 predictions to those of ETF-SI~(Q) in
Fig.~5(b). From the nuclei of Hf $(Z=72)$ to about Sm $(Z=62)$, there 
is not much difference between the results of ETF-SI and ETF-SI~(Q).
Therefore, for these nuclei, a comparison of NL-SV1 shell gaps with 
ETF-SI~(Q) ones is similar to that with ETF-SI as shown in Fig.~5(b).
In the region of r-process nuclei $Z=64-68$, the shell gaps between
NL-SV1 and ETF-SI~(Q) are similar. However, due to an extra quenching 
added in ETF-SI~(Q), differences in the shell gaps of the two 
approaches begin appearing below Z=62. The shell gap with ETF-SI~(Q)
becomes especially small for nuclei near the drip line. The impact of the 
variations in the shell effects in nuclei can not be visualized 
without comprehensive r-process calculations. In the case of 
ETF-SI~(Q), such calculations do exist and have shown promising 
results. On the other hand, r-process calculations using the 
results from NL-SV1 are being planned currently.

\subsection{Models with strong shell effects}

The strength of the shell effects at the major magic numbers has been
a point of numerous discussions in respect of nuclear abundances 
\cite{Kratz.93,Kratz.00,Pfeiffer.01}. The focus has mostly been 
on the possibilities and potential capabilities 
of various mass formulae of macroscopic-microscopic origin. This
is evidently due to the fact that mass formulae have been able to
\begin{figure*}
\centering
%\hspace{4cm}
\vspace{0.5cm}
\resizebox{0.60\textwidth}{!}{%
  \rotatebox{270}{\includegraphics{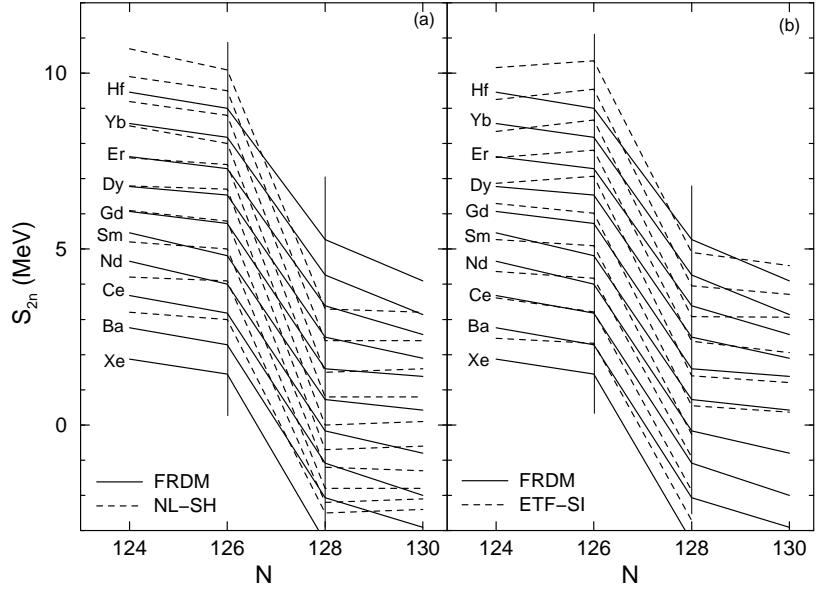}}
}
% If not, use
\vspace{0cm}       % Give the correct figure height in cm
\caption{The $S_{2n}$ values from the FRDM are compared to those (a) from 
the force with strong shell effects NL-SH and (b) the mass formula with
strong shell effects ETF-SI.}
\label{fig:6}       % Give a unique label
\end{figure*}
produce large-scale nuclear binding energies and other data relevant 
for use in network chain calculations. The microscopic theories have 
not yet enjoyed this privilege with the exception of isolated case(s)
where nuclear masses for the purpose have been calculated meaningfully.
Here, we wish to take stock of the situation on the shell effects of
nuclei within various frameworks.

We compare results of the mass formulae FRDM and ETF-SI, both of which 
show strong shell effects along with those from the RMF force NL-SH 
also exhibits strong shell effects. The $S_{2n}$ values from FRDM and 
NL-SH are shown across the shell closure $N=126$ in Fig.~6(a). 
The parallel lines for $S_{2n}$ with FRDM from Hf to Xe in the
the region of the shell gap bounded by the vertical lines in Fig.~6(a)
indicate the constancy of the shell gap with FRDM as discussed above.
The shell gap as indicated by the $S_{2n}$ values from NL-SH and the ensuing
large slope of the curve between $N=126$ and $N=128$ demonstrates 
overly strong shell effects with the force NL-SH. In comparison, the shell
gap with FRDM is smaller than that with NL-SH by about 2 MeV for Hf nucleus.
This difference between NL-SH and FRDM persists even for r-process nuclei 
in the region of $Z=62-66$. Thus, the shell effects at $N=126$ with NL-SH 
are significantly stronger than those with FRDM. However, due to natural 
decrease in the shell gap with an increase in isospin in the RMF theory, 
the shell gap with NL-SH does become comparable to FRDM for nuclei near 
the drip line. 

The indication that the shell strength with NL-SH is strong 
appeared in Ref. \cite{SLHR.94}, where the shell effects with NL-SH 
were studied across $N=82$. In that work, it was shown that the shell 
effects at $N=82$ with NL-SH were as strong as those with FRDM, but 
not stronger than FRDM. It was also shown that there was a remarkable 
agreement between the ground-state properties of the isotopic chains 
of Zr from $^{112}$Zr through to $^{130}$Zr in the two approaches.
Nuclei in this region exhibited not only similar values of quadrupole 
deformation, but also similar shape transitions across the region 
with FRDM and NL-SH \cite{SLHR.94}. This led to a surmise that the 
shell effects and shell structure with NL-SH are very similar to 
those of FRDM. However, this does not seem to be the case in the 
region of $N=126$. 

Taking into consideration the results of network chain calculations 
performed thus far, we believe that r-process network chain 
calculations with the force NL-SH or any similar microscopic force 
exhibiting strong shell effects are not expected to be successful,
at least for the third peak at $A \sim 190$.

We compare the shell effects with FRDM  to those with ETF-SI in 
Fig.~6(b). The shell gaps with ETF-SI for nuclei in the vicinity 
of Hf are more than 1 MeV larger than those with FRDM. The strong
shell gaps with ETF-SI are maintained across the r-process region 
and a near constancy of the shell gap at $N=126$ both with FRDM and
ETF-SI is seen clearly. The difference in the shell gap between 
ETF-SI and FRDM, however, decreases in going from the r-process path
nuclei towards the drip line nuclei. The fact remains that with 
ETF-SI the shell effects are stronger than those with FRDM. 

Using macroscopic-microscopic mass formulae FRDM and ETF-SI, 
it was shown \cite{Kratz.93,Thielemann.94,Kratz.95} that due to the 
stronger shell gaps in ETF-SI, r-process network chain calculations 
lead to a greater deficiency (trough) in the r-process nuclear 
abundances  below $A \sim 190$ peak than with FRDM. Thus, the earlier
conclusion of much of these analyses has been that  stronger shell 
effects at $N=126$ do not seem to be conducive to reproducing nuclear
abundances. However, it was shown in a later analysis 
\cite{ Thielemann.94} that in a ``realistic'' astrophysical scenario,
there is no stringent need for a quenching of the $N=126$ shell effects.
The trough that appeared near $A \sim 175$ in earlier analyses could be 
filled due to freeze-out effects even by using a mass model without quenching. 
This raises the possibility that under appropriate conditions, mass formulae
without a quenching of the shell strength at $N=126$ can be used
successfully.

\begin{figure}
\hspace{0cm}
\resizebox{0.45\textwidth}{!}{%
  \rotatebox{270}{\includegraphics{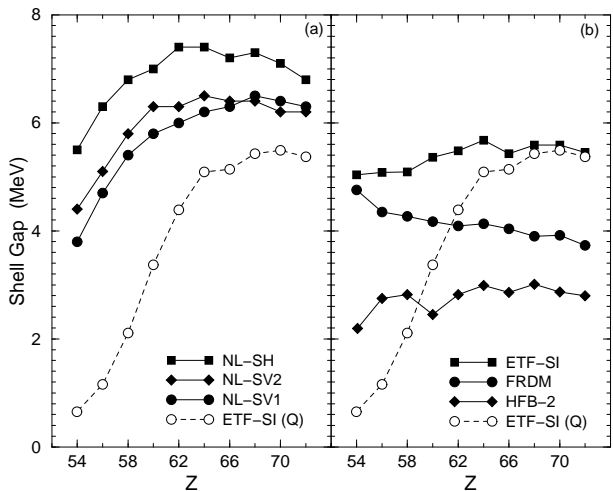}}
}
% If not, use
\vspace{0cm}       % Give the correct figure height in cm
\caption{The shell gaps $\Delta_S$ (Eq.~\ref{gap}) at $N=126$ 
as obtained from the RMF forces (a) NL-SH, NL-SV1 and NL-SV2 and 
compared with those from ETF-SI~(Q). (b) The shell gaps from mass 
formulae ETF-SI, FRDM and HFB-2 are shown. A comparison is also made 
with the ETF-SI~(Q).}
\label{fig:7}       % Give a unique label
\end{figure}

\subsection{The $N=126$ shell gap in nuclei: RMF versus mass formulae}

The status of $N=126$ shell gaps $\Delta_S$ as defined in Eq.~\ref{gap} is 
summarized in Fig.~7 for the RMF theory and for various mass formulae. 
The shell gap at $N=126$ as obtained from the RMF calculations with NL-SH,
NL-SV1 and NL-SV2 are shown in Fig.~7(a). All the microscopic
calculations show a decreasing shell gap in going through
the r-process region. This reduction is, however, a little stronger 
as one moves towards the drip line.

A comparison shows that NL-SH exhibits shell gaps which are 
consistently larger than those with NL-SV1 and NL-SV2. This characterizes
the stronger shell effects of NL-SH vis-a-vis other forces. The shell gaps
with NL-SV2 are slightly larger than those with NL-SV1. This was also
shown to be the case for shell gaps at $N=82$ with NL-SV2 \cite{Sharma.02}. 

We compare the shell gaps with the RMF forces to those from the quenched
mass formula ETF-SI~(Q) in Fig.~7(a). Evidently, the shell gaps with 
ETF-SI~(Q) are much smaller than those with NL-SH. However, these are also
smaller than those of the NL-SV1 and NL-SV2. The effect of the added 
quenching in ETF-SI~(Q) is apparent for nuclei below $Z=64$, 
where the shell gap is reduced significantly as compared to the nearly
constant values maintained in ETF-SI (see Fig.~7(b)). For nuclei 
below Nd $(Z=60)$, however, the quenching in the ETF-SI~(Q) is much 
stronger than the weakening of the shell effects predicted by NL-SV1. 

The shell gaps from various mass formulae are compared in Fig.~7(b).
Both ETF-SI and FRDM exhibit shell gaps which remain nearly constant
in going from the region of the r-process path towards the drip line.
Comparatively, ETF-SI shows shell effects that are stronger than
those with FRDM in much of the region shown in the figure. As shown in 
earlier calculations \cite{Pfeiffer.97}, the strength of the shell effects
with ETF-SI was found to produce much larger deficiency below 
$A \sim 190$ peak in the r-process abundances. 

The shell gaps with the new mass formula HFB-2 are compared with those
from ETF-SI and FRDM in Fig.~7(b). The shell gaps with HFB-2 are nearly 
constant as shown also by the other mass formulae. These are, however,
systematically smaller than those of ETF-SI and FRDM. There is a tendency 
of only a slight decrease in the shell gap in going towards the drip 
line nuclei. It is interesting to note that the addition of a 
Bogoliubov based pairing in the mass formula has not been found to 
be sufficient to suppress further the shell gaps below the r-process 
region. A comparison of the shell gaps from 
ETF-SI, FRDM and HFB-2 with those from ETF-SI~(Q) shows that 
the shell gaps from ETF-SI~(Q) are in striking contrast to all 
the other mass tables. However, as mentioned earlier, this striking 
difference has been brought about by the introduction of a quenching 
based upon results of HFB+SkP calculations. This extra weakening
of the shell gaps in the drip line region is not shown by any of 
the widely used mass formulae. It has been reported in some calculations
\cite{Pear.96, Kratz.95} that a weakening of the shell effects 
in the r-process region to the drip line is required 
for reproducing the r-process abundances around the peaks at 
$A \sim 130 $ and $A \sim 190$. 

\subsection{The shell $N=126$ versus $N=82$}

We discussed the evolution of the shell gap at $N=82$ near the
r-process path in the RMF theory in Ref. \cite{Sharma.02} in detail. 
The shell gaps at $N=82$ were found to weaken in the region of 
the r-process path. In going to the drip line, the shell effects
at $N=82$ showed a substantial reduction in the shell strength, eventually
leading to a complete disappearance of the shell gap for the drip line
nuclei.

\begin{figure}
\hspace{0cm}
\resizebox{0.45\textwidth}{!}{%
  \rotatebox{270}{\includegraphics{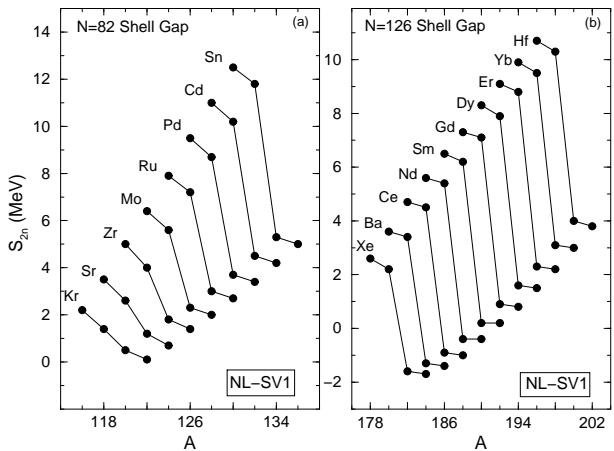}}
}
% If not, use
\vspace{0cm}       % Give the correct figure height in cm
\caption{The $S_{2n}$ values from NL-SV1 are shown for nuclei 
across the shell closure (a) $N=82$ and (b) $N=126$ in the 
region of the r-process path. The evolution of the shell 
gaps at $N=126$ is compared to that at $N=82$.}
\label{fig:8}       % Give a unique label
\end{figure}

We show in Fig.~8 a comparison of the shell effects at $N=126$ 
as obtained in the present work with those at $N=82$ \cite{Sharma.02}
in the region of the r-process path to the drip line. In order to 
visualize the shell gap at $N=82$, we show the $S_{2n}$ values 
obtained with NL-SV1 across the magic number for nuclides of
Sn $(Z=50)$ down to the drip line nuclides of Kr $(Z=36)$. 
The shell gap at $N=82$ shows a strong decrease in going from 
Cd $(Z=48)$ to Zr $(Z=40)$ in the region of the r-process path
as mentioned above. This reduction in the shell gap at $N=82$ is 
significantly faster than the corresponding shell gap at $N=126$ 
(Fig.~8(b)). In the latter case, the shell gap shows a much slower 
decrease in the region of the r-process nuclei from 
Er $(Z=68)$ to Nd $(Z=60)$. The $N=126$ shell strength exhibits 
a resilience to the change in isospin.

\subsection{The shell effects and r-process nucleosynthesis}

In the broader context of the shell effects and their implications on
the r-process nucleosynthesis, it is pertinent to discuss recent
r-process calculations which employ newly developed mass formulae
HFB-2 and HFB-7 \cite{Wanajo.04}. In this work, effect of various 
mass formulae on r-process nucleosynthesis has been studied using 
astrophysical model of prompt supernova explosion from a collapsing 
O-Ne-Mg core \cite{Wanajo.03}. In the mass formulae HFB-2 and HFB-7 
which have been used, pairing has been taken into account by the 
Bogoliubov method in a Skyrme density-functional approach. R-process
calculations in this study have also been carried out using the nuclear
masses from FRDM \cite{Moeller.95} and from the old droplet mass (DM)
model \cite{Hilf.76}.

It has been shown \cite{Wanajo.04} that due to the shell effects which 
are significantly reduced (weak) with HFB-2 and HFB-7, the abundance 
curve in the third peak at $A \sim 190$ is spread leading to a shift 
of the peak and consequently the valley at $A \sim 183$ is shifted 
significantly to lower masses. The results show that by the use of 
HFB-2 and HFB-7 masses, there are large deviations in the third peak as 
compared to the solar-system r-process abundances. Thus, due to the
weakness of the shell effects at $N=126$ near the r-process in the mass
models HFB-2 and HFB-7, there is a significant overproduction of nuclei
to the left of the third peak in the solar pattern.
It is interesting to note that on using masses due to FRDM and DM,
which are characterized by stronger shell effects at $N=126$, it has
been shown \cite{Wanajo.04} that the abundance curve gives rise 
to a sharp r-process peak at $A\sim 190$, in better agreement 
with the solar pattern in the region. These results suggest that
in the prompt supernova model considered in Ref. \cite{Wanajo.04}
and conditions applicable therein, mass models/microscopic theories
with stronger shell effects at $N=126$ would reproduce 
the main features of the solar 
r-process abundance curve around the third peak reasonably well. 
It may be recalled that rather similar conclusion was reached in Ref.
\cite{Thielemann.94}, wherein it was shown that a mass model
without quenching at $N=126$ can fill up the trough at $A \sim 175$
and reproduce the abundance curve near the third peak due to
freeze-out effects.

In view of the different behaviour of the shell effects at $N=82$
and $N=126$ as predicted by NL-SV1 in the RMF theory as shown in 
Fig.~8, it is pertinent to discuss the results of the r-process 
calculations of Ref. \cite{Wanajo.04} around the second peak at 
$A \sim 130$. The results have shown that droplet mass models FDRM
and DM with strong shell effects also at $N=82$ produce troughs 
(underproduction) at $A \sim 115$ and $A \sim 140$ in the abundance
curve. This shows that in contrast with $N=126$, stronger shell effects 
at $N=82$ are not desirable for reproducing the solar abundance pattern.
On the other hand, with HFB-2 and HFB-7 deficiencies in the
r-process production below and above the second peak at $A \sim 130$
are significantly remedied especially with HFB-2. However, the weak
shell effects inherent in HFB-2 and HFB-7 have the consequence that 
abundances around $A \sim 130$ are also spread out as opposed to the
solar system r-process abundances. 

The results of Ref. \cite{Wanajo.04} indicate that within the model 
employed in this work, the two peaks at $A \sim 130$ and 
$A \sim 190$ require a different nature of the shell effects. For a
successful reproduction of abundances near $A \sim 130$, shell effects
at $N=82$ in the r-process region should not be strong, whereas
the third peak seems to require moderate to strong shell effects
at $N=126$ in contrast to $N=82$. Thus, in the model considered in
Ref. \cite{Wanajo.04}, the analysis of the second and 
the third peaks in the solar-system r-process abundance 
curve suggests a different nature of the shell effects at $N=82$ 
and at $N=126$.

Microscopic RHB calculations with the Lagrangian model NL-SV1 show
two very different features for the shell effects at $N=82$ and 
$N=126$, as depicted in Fig.~8. These features seem to be consistent
with the picture that emerges from the results of Ref. \cite{Wanajo.04}.  
It remains to be seen in future network chain calculations with 
masses obtained from microscopic calculations with NL-SV1 whether
the two different features of the shell effects exhibited 
by NL-SV1 would suffice towards reproducing the r-process 
abundances near the second and the third peak.

\section{Conclusions}

We have investigated the shell effects in nuclei at the magic 
neutron number $N=126$ in the region of the r-process 
path using the relativistic Hartree-Bogoliubov approach within
the relativistic mean-field theory. Two Lagrangian models, one with 
the nonlinear scalar coupling of $\sigma$ meson and another one 
that includes a nonlinear vector coupling of $\omega$ meson have 
been considered. A comparison of the shell effects in the RMF theory
has been made with the predictions of various mass formulae. It is shown 
that the predictions of RMF calculations exhibit a slight reduction 
of the shell gap in going from the r-process path to the neutron
drip line irrespective of the shell strength exhibited by a Lagrangian
parameter set along the stability line. This is slightly different
from a near constancy of the shell gaps demonstrated by major mass formulae 
in the region of the r-process path and the drip line. Consequences of 
the shell strength on the r-process nucleosynthesis have been discussed.

It is shown that the Lagrangian force NL-SV1 with the vector 
self-coupling of $\omega$ meson, which reproduces the shell gaps 
along the stability line, shows that the shell effects at $N=126$ 
exhibit only a marginal reduction in the shell strength in going from 
the r-process path to the drip line. Consequenty, the shell 
effects retain a strong character along the r-process path 
at $N=126$. This is in striking contrast to the earlier
results \cite{Sharma.02} with NL-SV1 that in the RMF theory 
shell effects at $N=82$ exhibit a significant weakening of the strength
in going from the r-process path to the neutron drip line. This shows 
that different magic numbers may exhibit a different nature of the 
shell effects in the extreme regions of the periodic table.
 
Analysis of the results of a recent r-process calculations \cite{Wanajo.04} 
has suggested that stronger shell effects at $N=126$ and comparatively 
weaker shell effects at $N=82$ are conducive to reproducing 
the r-process abundances in the second and the third peak, respectively,
in the solar-system r-process abundance curve. Our results exhibit
features which are consistent with this analysis and support the
conjecture that a different nature of the shell effects may be at 
play in r-process nucleosynthesis of heavy nuclei.

A scrutiny of the shell effects with various RMF forces and with
various mass formulae has shown that a given nature of the shell 
effects in a known region may not be sufficient to calibrate the 
shell structure in unknown regions. Accordingly, an extrapolation in 
the unknown regions of the r-process path is frought with an 
uncertainty.

\begin{acknowledgments}
This work is supported by the Research project No. SP01/02 of 
the Research Administration, Kuwait University.
\end{acknowledgments}

\end{document}